# New Approaches in Synthetic Biology: Abiotic Organelles and Artificial Cells Powered and Controlled by Light

Günther Knör

Johannes Kepler University Linz, Institute of Inorganic Chemistry – Center for Nanobionics and Photochemical Sciences (CNPS), Altenberger Strasse 69, 4040 Linz, Austria

**Abstract.** One of the major goals of nanobionics and synthetic biology is the development of artificial cell organelles for the creation of cell-like structures operating similar to biological systems with a minimalistic set of building blocks. To achieve this ambitious goal, two major design strategies are followed in the field of synthetic biology. The top-down approach tries to generate a radically trimmed but still intact artificial cell by eliminating all non-essential components from the much more complex native systems. In contrast, bottom-up synthetic biology aims at constructing a functioning minimal cell by combining together all essential building blocks step-by-step starting from scratch.
In the present contribution, the author´s ongoing activities to develop artificial reaction centers for novel types of photoautotrophic processes and to provide fully biocompatible synthetic enzyme counterparts and artificial organelles as abiotic building blocks for bottom-up synthetic biology are summarized. This unique approach has the potential to create unprecedented minimal artificial cell-like systems that can be powered and readily controlled by light as an external stimulus.

**Keywords:**, Nanobionics, Photochemistry, Artificial Enzymes, Light-controlled Metabolism, Bottom-up Synthetic Biology, Abiotic Cells, Synthetic Organelles, Chemical-Biological Hybrid Systems, Photoautotrophic Biotechnology.

## 1 Introduction

In current transdisciplinary research many efforts are made to better understand the boundary conditions and minimal requirements connected to the origin, distribution, technical exploitation and sustain of living systems [1]. Scientists active in the field of synthetic biology [2] systematically explore the fundamental principles and limiting constraints existing for all organisms by creating non-native counterparts of biological systems. One of the ambitious goals of many researchers active in the field is trying to construct a primitive minimal version of a functioning artificial cell. This occurs either by top-down strategies, where apparently non-essential components of intact organisms are removed until a new kind of artificial cellular structure remains [3], or in a bottom-up fashion, where all required components are combined together step-by-step [4]. These two complementary approaches are schematically illustrated in Figure 1.

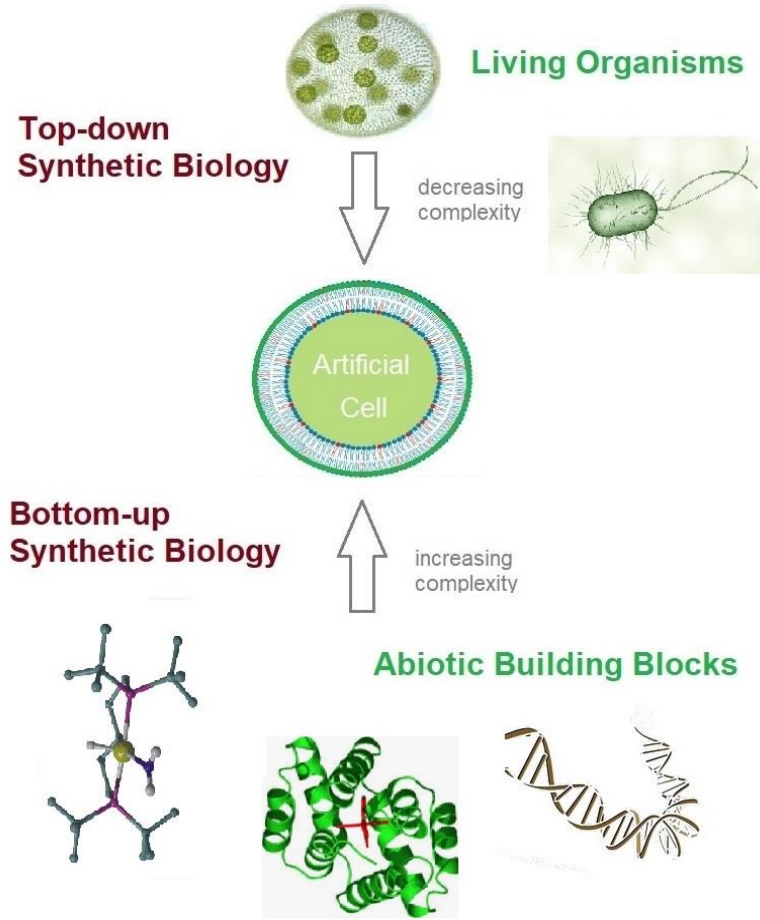

**Fig. 1.** Comparison of the two different strategies followed in current synthetic biology research with the common goal of generating a minimal version of an artificial cell (center). Elimination of all non-essential components from native cells in order to reduce the level of complexity of living systems represents the so-called "top-down" approaches (top). The "bottom-up" approach aims at the creation of a artificial cell-like system by bringing together a minimal set of functional synthetic building blocks (bottom).

Up to now, however, all claimed examples of top-down designed artificial cells still have to rely completely on the blueprints of naturally evolved structures and on the other hand no functioning cell-like constructs based on purely abiotic building blocks could be obtained with conventional bottom-up synthetic biology strategies. In the following sections, an alternative and now gradually maturing approach to achieve a minimal artificial cell based on light-controlled building blocks is briefly introduced.

## 2   Photochemical control of Intracellular Functions

### 2.1   Light-mediated energy supply and information flow

Life requires an orchestrated flow of energy, matter and information. Photons are ideally suited to target for the construction of a minimal cell-like structure, since they can selectively interact with matter as the fundamental carriers of energy and information with an excellent spatial and temporal resolution at the speed of light [7]. Such an approach has been systematically elaborated in the last decades [5,6] and now allows to switch on and off the desired functions inside both native and artificial cells on demand [7]. At the same time, the application of synthetic building blocks for the photochemical control of intracellular functions represents the most straightforward strategy for creating independent photoautotrophic systems including artificial photosynthetic reaction centers [8] that can be directly supplied with solar radiation as their only energy source for metabolism.

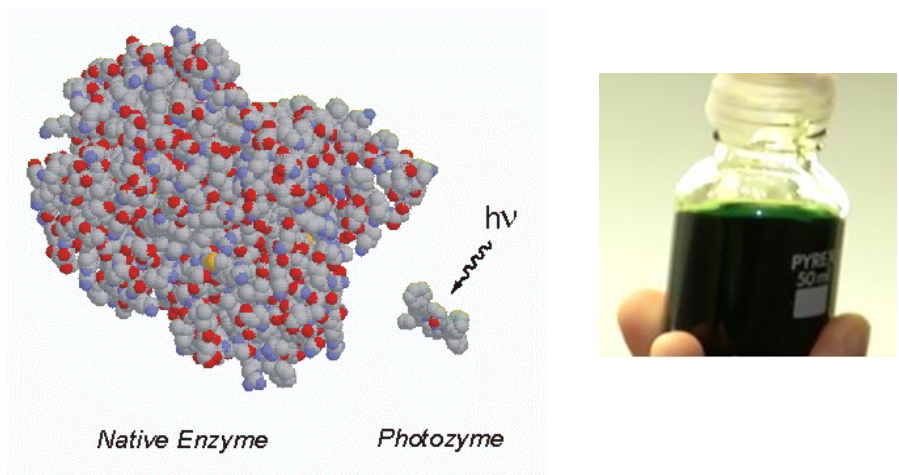

**Fig. 2.** Schematic representation of the strategy to substitute the functional role of protein- and nucleic acid-based native counterparts by low-molecular-weight light-responsive key-components for creating minimalistic versions of photochemically controlled artificial organelles and cells [5-7].

## 2.2 Light-triggered metabolic pathways

One of the logical expansions of this versatile concept requires the control of photocatalytic one-pot multistep reactions using different excitation wavelengths or other orthogonal tools to address a certain set of photocatalysts for driving vectorial substrate conversion cascades resembling natural metabolic pathways. Once the functioning light-controlled components replacing native biological key-steps have been designed, their coupling to cascade processes in a metabolism-like fashion indeed also becomes possible [7]. Controlled and switched on and off by means of light as the external stimulus, this stepwise biomimetic transformation of substrates and metabolites (A→B→C→…) can occur in a one-pot version in the same cell-like compartment [7] or could also be driven across self-assembled sub-structures separated by permeable membranes as schematically depicted in Figure 4.

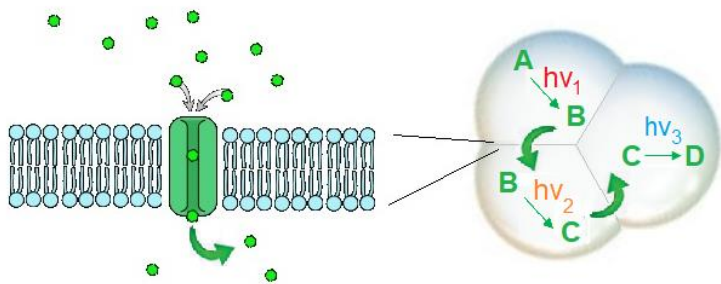

**Fig. 3.** Stepwise metabolism-like substrate conversion fully controlled by light excitation using different wavelength and intensity as an external regulation mechanism. The individual photocatalytic steps are carried out in self-assembled vesicle structure with membrane pores allowing an exchange of the metabolites similar to the model systems introduced by Elani *et al.* [9].

In the context of artificial enzyme function, meanwhile, the rational application of photochemical key-steps in such biomimetic catalytic systems has been systematically elaborated in depth, and it could be shown that competitive or even superior functional counterparts of native oxidoreductase enzymes, nucleases and even more complex multienzyme reaction centers are readily obtained with this new strategy based on photoassisted key-steps and full spatio-temporal light-control of catalytic performance [7]. Regarding the bioenergetic aspects of this approach, the first fully functional counterpart of the complete energy-trapping and solar-to-fuel conversion cascade otherwise only feasible with the natural photosystem I of green plants has been reported [8].
Another step forward to mimicking natural systems of higher complexity is the incorporation of artificial photoenzymes into chemical environments such as artificial cell-like structures, tissue-like environments or synthetic organelles to provide defined three-dimensional architectures with a given light-controllable function. First steps in this direction are discussed in the next section below.

## 3      Carrier Matrices for Artificial Organelles and Cells

In nature all living cells consist of a cell membrane enclosing the water-based cytoplasma carrying a plethora of different functional components and metabolites. In larger biological tissues, ensembles of many similar cells are connected together to achieve their specific functions in a synergetic way. The design of a primitive synthetic carrier material for artificial cells could thus include the shape-defining properties of an abiotic tissue matrix permeated by the cytosol as the reaction medium and containing cell-like compartments and embedded or attached artificial organelles as the functional subunits. Soft and optically transparent hydrogel materials represent a straightforward option for the verification of such an artificial three-dimensional tissue matrix for bottom-up synthetic biology applications due to their excellent biocompatibility, high water content and well-established usefulness in the fields of life-sciences and medicine (Figure 4).

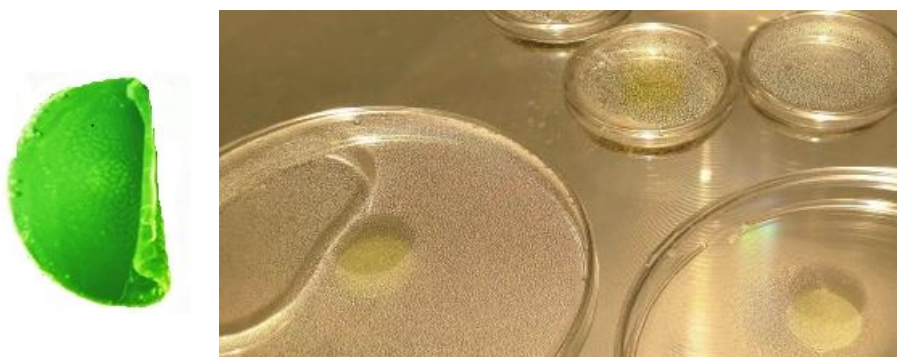

**Fig. 4.** Photoenzyme model compounds and artificial photosynthetic reaction centers from the author´s lab tested for their performance in cell and tissue-like hydrogel environments.

Although these latter aspects are currently still in their infancy and a functioning artificial cell is not in sight today, definitely a bright future can be expected for the application of light-control and photochemical key-steps in the field of bottom-up biology.

### Acknowledgement

Support by the Austrian Science Foundation (FWF Project W1250 DK9: "Photochemical Control of Cellular Functions") is gratefully acknowledged.